\documentclass[conference]{IEEEtran}

\usepackage{graphicx}
\usepackage{graphics}
\usepackage{color}
\usepackage{psfrag}
\usepackage{pstricks, pst-node, pst-tree}
\usepackage{pstricks-add}
\usepackage{latexsym}
\usepackage{mdwmath}
\usepackage{amsmath}
\usepackage{amssymb}

\usepackage{color}
\usepackage{psfrag}
\usepackage{alltt}
\usepackage{moreverb}

\usepackage{graphicx}
\usepackage{psfrag}
\usepackage{epsfig}
\usepackage{enumerate}
\usepackage{amssymb}
\usepackage{times}

\usepackage[pdfborder={0 0 0 },bookmarks=false,draft=true]{hyperref}
\usepackage{ifthen}

\newtheorem{theorem}{Theorem}

\newtheorem{proposition}[theorem]{\bf{Proposition}\\}

 \makeatletter
\@addtoreset{propositionpart}{proposition}
\makeatother

\newcounter{theillustrationexample}
\definecolor{lightgray}{RGB}{230,230,230}

\newcommand{\myvec}[1]{ \mathbf{#1} }
\newcommand{\myvecsymb}[1]{ \boldsymbol{#1} }
\newcommand{\transp}[1]{ {#1}^T }
\newcommand{\mymatrix}[1]{ \mathbf{#1} }

\newcommand{\R}[1]{ \mathbb{R}^{#1}}

\newcommand{\mathnorm}[1]{ \|{#1}\|}

\newcommand{\set}[1]{\mathit{#1}}
\newcommand{\grandset}[1]{\mbox{$\cal{#1}$}}

\newcommand{\bigonotation}[1]{\mbox{$\cal{O}$}(#1)}
\newcommand{\labelfunc}[1]{\text{\it I}(#1)}

\newcommand{\hypersurface}[3]{S\left({\frac{#2}{|{#2}|}}{#1}_{|{#2}|,:}\,;\, {\frac{#3}{|{#3}|}}{#1}_{|{#3}|,:}\right)}

\long\def\symbolfootnote[#1]#2{\begingroup%
\def\thefootnote{\fnsymbol{footnote}}\footnote[#1]{#2}\endgroup}

\begin{document}


\def\thepropositionpart{\alph{propositionpart}}

\author{\IEEEauthorblockN{Megasthenis Asteris and Dimitris S. Papailiopoulos}
\IEEEauthorblockA{Department of Electrical Engineering\\
University of Southern California\\
Los Angeles, CA 90089 USA\\
Email: \texttt{\{asteris, papailio\}@usc.edu}}
\and
\IEEEauthorblockN{George N. Karystinos}
\IEEEauthorblockA{Department of Electronic and Computer Engineering\\
Technical University of Crete\\
Chania, 73100, Greece\\
Email: \texttt{karystinos@telecom.tuc.gr}}
}

\title{Sparse Principal Component of a Rank-deficient Matrix}
\maketitle

\begin{abstract}
We consider the problem of identifying the sparse principal component of a rank-deficient matrix.
We introduce auxiliary spherical variables and prove that there exists a set of candidate index-sets (that is, sets of indices to the nonzero elements of the vector argument) whose size is polynomially bounded, in terms of rank, and contains the optimal index-set, i.e. the index-set of the nonzero elements of the optimal solution.
Finally, we develop an algorithm that computes the optimal sparse principal component in polynomial time for any sparsity degree.
\end{abstract}

\section{Introduction}
Principal component analysis (PCA) is a well studied, popular tool used for dimensionality reduction and low-dimensional representation of  data with applications spanning many fields of science and engineering.
Principal components (PC) of a set of ``observations'' on some $N$ variables capture orthogonal directions of maximum variance and offer a Euclidean-distance-optimal, low-dimensional visualization that -for many purposes- conveys  sufficient amount of information.
Without additional constraints, the PCs of a data set can be computed in polynomial time in $N$, using the eigenvalue decomposition.

One disadvantage of the classical PCA is that, in general, the extracted eigenvectors are expected to have nonzero elements in all their entries.
However, in many applications sparse vectors that maximize variance are more favorable.
Sparse PCs can be less complicated to interpret, easier to compress, and cheaper to store.
Thus, if the application requires it, then some of the maximum variance property of a PC may be slightly traded for sparsity.
To mitigate the fact that PCA is oblivious to {\it sparsity} requirements, an additional cardinality constraint needs to be introduced to the initial variance maximization objective.
The sparsity aware flavor of PCA, termed {\it sparse PCA}, inarguably comes at a higher cost: sparse PCA is an NP-Hard problem \cite{moghaddam_weiss_avidan_2006a}.

To approximate sparse PCA various methods have been introduced in the literature.
Initially, factor rotation techniques that extract sparse PCs were used in \cite{daiser_1958}, \cite{jolliffe_1995}.
Straightforward thresholding of PCs was presented in \cite{cadima_jolliffe_1995} as a computationally light means for obtaining sparsity.
Then, a modified PCA technique based on the LASSO was introduced in \cite{jolliffe_2003}.
In \cite{zou_hastie_tibshirani_2006} an elaborate nonconvex regression-type optimization approach combined with LASSO penalty was used to approximately tackle the problem.
A nonconvex technique, locally solving difference-of-convex-functions programs was presented in \cite{sriperumbudur_torres_lanckriet_2007}.
Semidefinite programming (SDP) was used in \cite{daspremont_chaoui_jordan_lanckriet_2007}, \cite{zhang_daspermont_elghaoui_2010}, while \cite{daspermont_bach_elghaoui_2008} augmented the SDP approach with an extra greedy step that offers favorable optimality guarantees under certain sufficient conditions.
The authors of \cite{moghaddam_weiss_avidan_2006b} considered greedy and branch-and-bound approaches, further explored in \cite{moghaddam_weiss_avidan_2007}.
Generalized power methods using convex programs were also used to approximately solve sparse PCA \cite{journee_nesterov_richtarik_sepulchre_2008}.
A sparse-adjusted deflation procedure was introduced in \cite{mackey_2009} and in \cite{amini_wainwright_2009} optimality guarantees were shown for specific types of covariance matrices under thresholding and SDP relaxations.

{\bf Our Contribution:}
In this work we prove that the sparse principal component of a matrix ${\bf C}$ can be obtained in polynomial time under a new sufficient condition: when  ${\bf C}$ can be written as a sum of a scaled identity matrix plus an update, i.e. ${\bf C}=\sigma{\bf I}_N+{\bf A}$, and the rank of the update ${\bf A}$ is not a function of the problem size.\footnote{If $\sigma=0$, then we simply have a low-rank matrix ${\bf C}$.}
Under this condition, we show that sparse PCA is polynomially solvable, with the exponent being only a linear function of the rank.
This result is possible after introducing {\it auxiliary spherical variables} that ``unlock'' the low-rank structure of ${\bf A}$.
The low-rank property along with the auxiliary variables enable us to scan a constant dimensional space and identify a polynomial number of {\it candidate} sparse vectors.
Interestingly, we can show that the optimal vector always lies among these candidates and a polynomial time search can always retrieve it.

\section{Problem Statement}
We are interested in the computation of the real, unit-length, and at most $K$-sparse principal component of the $N\times N$ nonnegative definite matrix ${\bf C}$, i.e.
\begin{equation}
 \myvec{x}_{\text{opt}} \stackrel{\bigtriangleup}{=} \arg\max_{{\bf x}\in \mathbb{S}_K^N} {\bf x}^T{\bf C}{\bf x}
 \label{initial_problem}
\end{equation}
where  $\mathbb{S}_K^N \stackrel{\bigtriangleup}{=}\left\{{\bf x}\in \R{N}:  \| \myvec{x}\| =1,\; \text{card}({\myvec{x}}) \le K \right\}$.
Interestingly, when ${\bf C}$ can be decomposed as a low-rank update of a constant identity matrix, i.e. ${\bf C}=\sigma{\bf I}_N+{\bf A}$ where $\sigma\in\mathbb{R}$, ${\bf I}_N$ is the $N\times N$ identity, and ${\bf A}$ is a nonnegative definite matrix with rank $D$, then ${\bf x}^T{\bf C}{\bf x}=\sigma\|{\bf x}\|^2+{\bf x}^T{\bf A}{\bf x}$.
Therefore, the optimization (\ref{initial_problem}) can always be rewritten as
\begin{equation}
\myvec{x}_{\text{opt}} = \arg\max_{{\bf x}\in \mathbb{S}_K^N} {\bf x}^T{\bf A}{\bf x}
\end{equation}
where the new matrix ${\bf A}$ has rank $D$.
Since ${\bf A}$ is a nonnegative definite matrix, it can be decomposed as $\mymatrix{A} = \mymatrix{V}\transp{\mymatrix{V}}$ where
$\mymatrix{V} \stackrel{\bigtriangleup}{=} [\myvec{v}_1 \hspace{.2cm} \myvec{v}_2 \hspace{.2cm} \cdots \hspace{.2cm} \myvec{v}_D]$
and problem \eqref{initial_problem} can be written as
\begin{equation}
 \myvec{x}_{\text{opt}}= \arg\max_{{\bf x}\in \mathbb{S}_K^N}  {\transp{\myvec{x}}\mymatrix{V}\transp{\mymatrix{V}}\myvec{x}} = \arg\max_{{\bf x}\in \mathbb{S}_K^N}  \left\|\transp{\mymatrix{V}}\myvec{x}\right\|.
 \label{simpler_problem}
\end{equation}
In the following, we show that when $D$ is {\it not} a function of $N$, (\ref{initial_problem}) can be solved in time $\mathcal{O}\left(\text{poly}(N)\right)$.

\section{Rank-$1$ and Rank-$2$ Optimal Solutions}
Prior to presenting the main result for the general rank $D$ case, in this section we provide insights as to why sparse PCA of rank deficient matrices can be solved in polynomial time, along with the first nontrivial case of polynomial solvability for rank-$2$ matrices ${\bf A}$.
\subsection{Rank-$1$: A motivating example} \label{section:rank_1_case}
In this case, $\mymatrix{A}$ has rank $1$, $\mymatrix{V}=\myvec{v}$, and (\ref{simpler_problem}) becomes
\begin{flalign}
\myvec{x}_{\text{opt}} 
& =  \arg\max_{{\bf x}\in \mathbb{S}_K^N} \left|{\bf v}^T{\bf x}\right|= \arg\max_{{\bf x}\in \mathbb{S}_K^N}\left|\sum_{i=1}^Nv_ix_i\right|.\label{rank1maxQF}
\end{flalign}
It is trivial to observe that maximizing (\ref{rank1maxQF}) can be done by distributing the $K$ nonzero {\it loadings} of  ${\bf x}$ to the $K$ absolutely largest values of ${\bf v}$, indexed by set $\set{I}$.
Then, the optimal solution $\myvec{x}_{\text{opt}}$ has the following $K$ nonzero loadings
\begin{equation}
{\bf x}_{\text{opt},\set{I}} = \frac{{\bf v}_{\set{I}}}{\left\|{\bf v}_{\set{I}}\right\|}.
\end{equation}
where ${\bf x}_{\text{opt},\set{I}}$ denotes the set of elements of ${\bf x}_{\text{opt}}$ indexed by $\set{I}$ and $x_{\text{opt},i} = 0$ for all $i\notin \set{I}$.

The leading complexity term of this solution is determined by the search for the $K$ largest element of ${\bf v}$, which can be done in time $\mathcal{O}(N)$ \cite{cormen}.
Therefore, the rank-$1$-optimal solution can be attained in time that is linear in $N$.

\subsection{Rank-2: Introducing spherical variables}

In this case, $\mymatrix{V}$ is an $N\times 2$ matrix.
A key tool used in our subsequent developments is a set of auxiliary spherical variables.
For the rank-$2$ case, we introduce a single phase variable $\phi \in \left(-\frac{\pi}{2},\; \frac{\pi}{2}\right]$ and define the polar vector
\begin{equation}
 \myvec{c}(\phi)  \stackrel{\bigtriangleup}{=} \begin{bmatrix}
							    \sin\phi\\
							    \cos\phi
                               \end{bmatrix}
\end{equation}
which lies on the surface of a radius $1$ circle.
Then, from Cauchy-Schwartz Inequality we obtain
\begin{flalign}
 \left|\myvec{c}^T(\phi)\transp{\mymatrix{V}}\myvec{x}\right| &\le \mathnorm{\myvec{c}(\phi)}\mathnorm{\transp{\mymatrix{V}}\myvec{x}}
 = \mathnorm{\transp{\mymatrix{V}}\myvec{x}} \label{rank2intro:cauchy_for_my_case}
\end{flalign}
with equality if and only if ${\bf c}(\phi)$ is parallel to ${\bf V}^T{\bf x}$.
Therefore, finding $\myvec{x}$ that maximizes $\mathnorm{\transp{\mymatrix{V}}\myvec{x}}$ in (\ref{simpler_problem}) is equivalent to obtaining ${\bf x}$ of the $(\myvec{x}, \phi)$ pair that maximizes $\left|\myvec{c}^T(\phi)\transp{\mymatrix{V}}\myvec{x}\right|$. 
Initially, this rewriting of the problem might seem unmotivated, however in the following we show that the use of ${\bf c}(\phi)$ unlocks the low-rank structure of ${\bf V}$ and allows us to compute the sparse PC of ${\bf A}$ by solving a polynomial number of rank-$1$ instances of the problem.

We continue by restating the problem in (\ref{simpler_problem}) as
\begin{flalign}
\max_{{\bf x}\in \mathbb{S}_K^N}{\mathnorm{\transp{\mymatrix{V}}\myvec{x}} }
&=
\max_{{\bf x}\in \mathbb{S}_K^N}
\max_{\phi \in (-\frac{\pi}{2},\, \frac{\pi}{2}]}
\left|\myvec{c}^T(\phi)\transp{\mymatrix{V}}\myvec{x}\right|\nonumber\\
&=\max_{\phi \in (-\frac{\pi}{2},\, \frac{\pi}{2}]}
\max_{{\bf x}\in \mathbb{S}_K^N} 
|\underbrace{\myvec{c}^T(\phi)\transp{\mymatrix{V}}}_{\myvec{v}^T(\phi)}\myvec{x}|
\label{rank2intro:less_naive_equivalent}.
\end{flalign}
If we fix  $\phi$, then the internal maximization problem 
\begin{flalign}
\max_{{\bf x}\in \mathbb{S}_K^N}
\left|\myvec{v}^T(\phi){\bf x}\right| \label{rank2intro:observation_for_given_phi_is_rank_1}
\end{flalign}
is a rank-$1$ instance, for which we can determine in time $\mathcal{O}(N)$ the optimal index-set $\set{I}(\phi)$ corresponding to the indices of the $K$ absolutely largest elements of $\myvec{v}(\phi) = \mymatrix{V} \, \myvec{c}(\phi)$.

However, why should $\phi$ simplify the computation of a solution?
The intuition behind the polar vector concept is that every element of
\begin{flalign}
 \mymatrix{V} \myvec{c}(\phi)
=
\begin{pmatrix}
	\displaystyle  V_{1,1}\sin{\phi} +  V_{1, 2}\cos{\phi}\\
		      \vdots \\
    \displaystyle  V_{N,1}\sin{\phi} +  V_{N, 2}\cos{\phi}\\
\end{pmatrix}\label{rank2intro:Vc}
\end{flalign}
is actually a continuous function of $\phi$, i.e. a curve in $\phi$.
Hence, the $K$ absolutely largest elements of  $\mymatrix{V} \myvec{c}(\phi)$ at a given point $\phi$ are functions of $\phi$. 
Due to the continuity of the curves, we expect that the index set $\set{I}(\phi)$ will retain the same elements in an area ``around'' $\phi$. 
Therefore, we expect the formation of regions, or ``cells'' on the $\phi$ domain, within which the indices of the $K$ absolutely largest elements of ${\bf v}(\phi)$ remain unaltered. 
A sorting (i.e., an $\set{I}$-set) might change when the sorting of the amplitudes of two element in ${\bf v}(\phi)$ changes.
This occurs at points $\phi$, where these two absolute values become equal, that is, points where two curves intersect.
Finding all these {\it intersection points}, is sufficient to determine cells and construct all possible candidate $\set{I}$-sets.
Among all candidate $\set{I}$-sets, lies the set of indices corresponding to the optimal $K$-sparse PC.
Exhaustively checking the $\set{I}$-sets of all cells, suffices to retrieve the optimal.
Surprisingly, the number of these cells is exactly equal to number of possible intersections among the amplitudes of ${\bf v}(\phi)$, which is exactly equal to $2\binom{N}{2}=\mathcal{O}\left(N^2\right)$, counting all possible combinations of element pairs and sign changes.

\begin{figure}[t]
    \centering
    \includegraphics[width=0.5\textwidth]{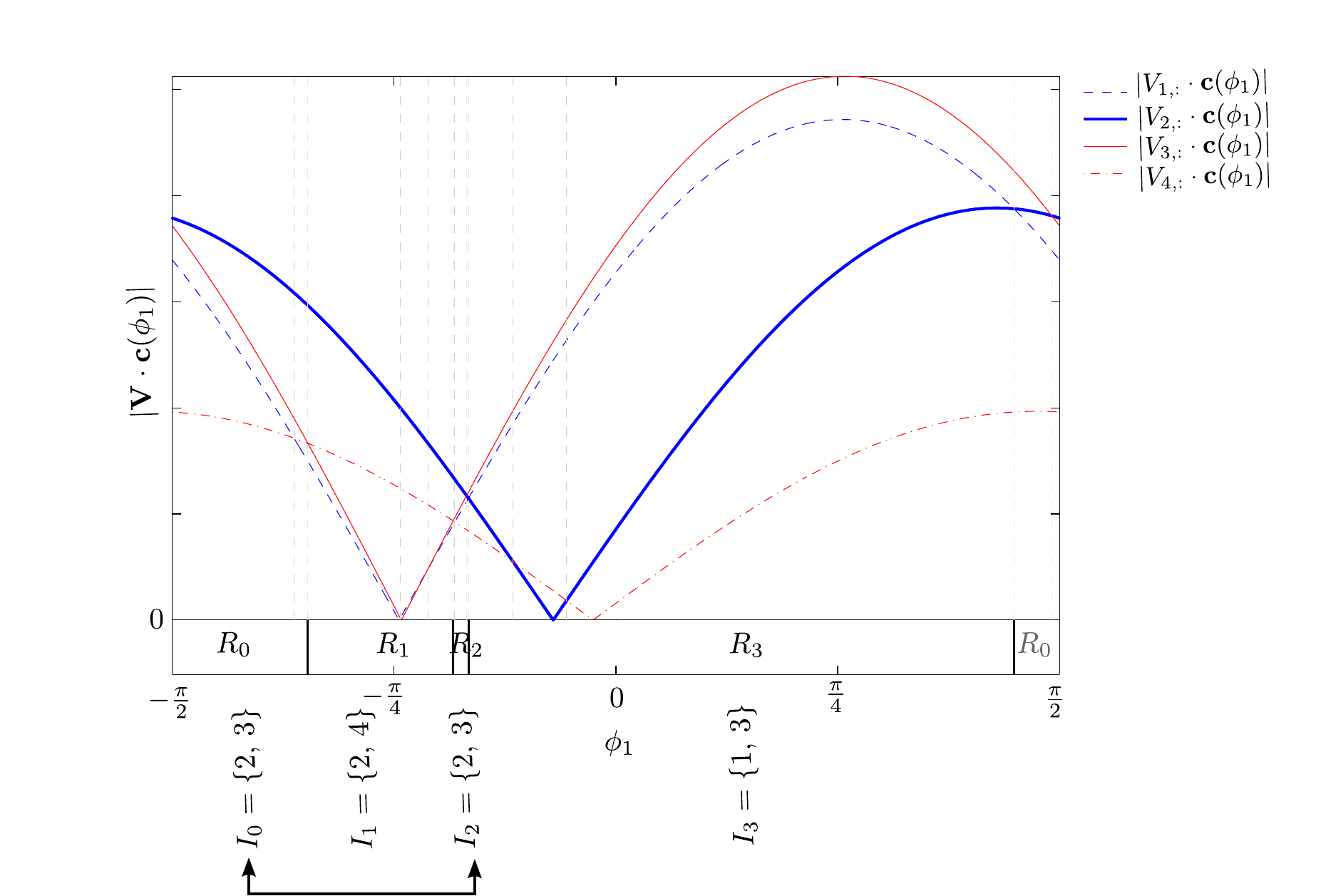}
    \caption{$N=4$, $K=2$, rank-$2$ case: Cells on the $\phi$ domain.}
    \label{rank2intro:fig:V_Rank2_Cells_and_Regions}
  \end{figure}
Before we proceed, in Fig. \ref{rank2intro:fig:V_Rank2_Cells_and_Regions} we illustrate the cell partitioning of the $\phi$ domain, where we set $N=4$ and $K=2$ and plot the magnitudes of the 4 curves that originate from the 4 rows of $\mymatrix{V}\myvec{c}(\phi)$. 
Cells (intervals) are formed, within which the sorting of the curves does not change. 
The borders of cells are denoted by vertical dashed lines at points of curve intersections. 
Our approach creates $12$ cells which exceeds the total number of possible index-sets, however this is not true for greater values of $N$. 
Moreover, we use $R_i$ regions to denote the sorting changes with respect to only the $K$-largest curves.
These regions is an interesting feature that might yet decrease the number of cells we need to check.
However, due to lack of space we are not exploiting this interesting feature here.

\subsection{Algorithmic Developments and Complexity}

Our goal is the construction of all possible candidate $K$-sparse vectors, determined by the index-sets of each cell on the $\phi$ domain.
This is a two step process. First we need to identify cell borders, and then we have to determine the index sets associated with these cells.

{\it Algorithmic Steps:}
We first determine all possible intersections of curve pairs in ${\bf v}(\phi)$.
Any pair  $\{i,j\}$ of distinct elements in ${\bf v}(\phi)$ is associated with two intersections: $v_i(\phi)=v_j(\phi)$ and $v_i(\phi)=-v_j(\phi)$. 
Solving these two equations with respect to $\phi$, determines a possible point where a new sorting of the $K$-largest values of ${\bf v}(\phi)$ might occur.

Observe that at an intersection point, the two values $v_i(\phi),v_j(\phi)$ are absolutely the same.
Exactly on the point of intersection, all but $2$ (the $i$th and $j$th) coordinates of a candidate $K$ sparse vector can be determined by solving a rank $1$ instance of the problem.
However, we are left with ambiguity with respect to $2$ coordinates $i$ and $j$ that needs to be resolved.
To resolve this ambiguity, we can visit the ``outermost'' point of the $\phi$ domain, that is, $\frac{\pi}{2}$.
There, due to the continuity of $v_i(\phi)$ and $\pm v_j(\phi)$, the sortings within the two cells defined by the two intersections, will be the identical, or opposite sortings of $v_i\left(\frac{\pi}{2}\right)$ and $\pm v_j\left(\frac{\pi}{2}\right)$, depending on whether $v_i(\phi)$ and $\pm v_j(\phi)$ are both positive, or negative at the intersection point, respectively.

Having described how to resolve ambiguities, we have fully described a way to calculate the $\set{I}$-set at any intersection point. 
Apparently, all intersection points, that is, all $\binom{N}{2}$ pairwise combinations of elements in ${\bf v}(\phi)$, have to be examined to yield a corresponding $\set{I}$-set.

{\it Computational Complexity:}
A single intersection point can be computed in time $\bigonotation{1}$.
At a point of intersection, we have to determine the $K$-th order element of the absolute values of ${\bf v}(\phi)$  and the $K-1$ elements larger than that, which can be done in time $\bigonotation{N}$. 
Resolving an ambiguity costs $\bigonotation{1}$.  
So in total, finding a single $\set{I}$-set costs $\bigonotation{N}$. 
Constructing all candidate $\set{I}$-sets requires examining all $2{N \choose 2}$ points, implying a total construction cost of $2{N \choose 2}\times \bigonotation{N}=\mathcal{O}\left(N^3\right)$.

\section{The Rank-$D$ Optimal Solution}
\label{section:maximization_with_new_method:subsection:theoretic_developments}
In the general case, $\mymatrix{V}$ is a $N\times D$ matrix.
In this section we present our main result where we prove that the problem of identifying the $K$-sparse principal component of a rank-$D$ matrix is polynomially solvable if the rank $D$ is not a function of $N$.
The statement is true for any value of $K$ (that is, even if $K$ is a function of $N$).
Our result is presented in the form of the following proposition.
The rest of the section contains a constructive proof of the proposition.

\begin{proposition}
Consider a $N\times N$ matrix ${\bf C}$ that can be written as a rank-$D$ update of the identity matrix, that is,
\begin{equation}
{\bf C}=\sigma{\bf I}_N+{\bf A}
\end{equation}
where $\sigma\in{\mathbb R}$ and ${\bf A}$ is a rank-$D$ symmetric positive semidefinite matrix.
Then, for any $K=1,2,\ldots,N$, the $K$-sparse principal component of ${\bf A}$ that maximizes
\begin{equation}
{\bf x}^T{\bf C}{\bf x}
\end{equation}
subject to the constraints $\| \myvec{x}\| =1$ and $\text{card}({\myvec{x}}) \le K$ can be obtained with complexity $\bigonotation{N^{D+1}}$.
\hfill$\Box$
\end{proposition}

We begin our constructive proof by introducing the spherical coordinates $\phi_1, \phi_2, \hdots, \phi_{D-1} \in (-\frac{\pi}{2},\; \frac{\pi}{2}]$ and defining the spherical coordinate vector
\begin{equation}
\myvecsymb{\phi}_{i:j} \stackrel{\vartriangle}{=} \transp{[\phi_{i}, \phi_{i+1},\hdots,\phi_{j}]},
\end{equation}
the hyperpolar vector
\begin{equation}
 \myvec{c}(\myvecsymb{\phi}_{1:D-1})  \stackrel{\bigtriangleup}{=} \begin{bmatrix}
							    \sin\phi_1\\
							    \cos\phi_1\sin\phi_2\\
							    \cos\phi_1\cos\phi_2\sin\phi_3\\
							    \vdots\\ 
							    \cos\phi_1\cos\phi_2\hdots\sin\phi_{D-1}\\
							    \cos\phi_1\cos\phi_2\hdots\cos\phi_{D-1}                                                         
                                                         \end{bmatrix},
\end{equation}
and set $\Phi\stackrel{\bigtriangleup}{=}\left(-\frac{\pi}{2},\frac{\pi}{2}\right]$.
Then, similarly with \eqref{rank2intro:less_naive_equivalent} and due to Cauchy-Schwartz Inequality, our optimization problem in (\ref{simpler_problem}) is restated as
\begin{equation}
\max_{{\bf x}\in \mathbb{S}_K^N} {\mathnorm{\transp{\mymatrix{V}}\myvec{x}} }=
\max_{{\bf x}\in \mathbb{S}_K^N}
\max_{\myvecsymb{\phi}_{1:D-1} \in \Phi^{D-1}}
\left|\transp{\myvec{x}} \mymatrix{V} \, \myvec{c}(\myvecsymb{\phi}_{1:D-1})\right| \label{naive_equivalent}.
\end{equation}
Hence, to find $\myvec{x}$ that maximizes $\mathnorm{\transp{\mymatrix{V}}\myvec{x}}$ in (\ref{simpler_problem}), we can equivalently find the $(\myvec{x}, \myvecsymb{\phi})$ pair that maximizes $\left|\transp{\myvec{x}}\mymatrix{V}\myvec{c}(\myvecsymb{\phi}_{1:D-1})\right|$.
We interchange the maximizations in $\eqref{naive_equivalent}$ and obtain
\begin{equation}
\max_{{\bf x}\in \mathbb{S}_K^N} {\mathnorm{\transp{\mymatrix{V}}\myvec{x}} }
=
\max_{\myvecsymb{\phi}_{1:D-1} \in \Phi^{D-1}}
\Big(
\max_{{\bf x}\in \mathbb{S}_K^N}
|\transp{\myvec{x}} \underbrace{\mymatrix{V} \, \myvec{c}(\myvecsymb{\phi}_{1:D-1})}_{\myvec{v}^{'}(\myvecsymb{\phi}_{1:D-1})}|
\Big).\label{less_naive_equivalent}
\end{equation}
For a given point $\myvecsymb{\phi}_{1:D-1}$, $\mymatrix{V} \, \myvec{c}(\myvecsymb{\phi}_{1:D-1})$ is a fixed vector and the internal maximization problem 
\begin{equation}
\max_{{\bf x}\in \mathbb{S}_K^N}
\left|\transp{\myvec{x}}\mymatrix{V} \, \myvec{c}(\myvecsymb{\phi}_{1:D-1})\right|
=
\max_{{\bf x}\in \mathbb{S}_K^N}
\left|\transp{\myvec{x}} \myvec{v}^{'}(\myvecsymb{\phi}_{1:D-1})\right|
\label{observation_for_given_phi_is_rank_1}
\end{equation}
is a rank-$1$ instance.
That is, for any given point $\myvecsymb{\phi}_{1:D-1}$, we can determine the optimal set $\set{I}(\myvecsymb{\phi}_{1:D-1})$ of the nonzero elements of $\myvec{x}$ as the set of the indices of the $K$ largest elements of vector $|\mymatrix{V} \, \myvec{c}(\myvecsymb{\phi}_{1:D-1})|$.

To gain some intuition into the purpose of inserting the second variable $\myvecsymb{\phi}_{1:D-1}$, notice that every element of $\pm \mymatrix{V} \myvec{c}(\myvecsymb{\phi}_{1:D-1})$ is actually a continuous function of $\myvecsymb{\phi}_{1:D-1}$, a $D$-dimensional hypersurface and so are the elements of $|\mymatrix{V} \myvec{c}(\myvecsymb{\phi}_{1:D-1})|$.
When we sort the elements of  $|\mymatrix{V} \myvec{c}(\myvecsymb{\phi}_{1:D-1})|$ at a given point $\myvecsymb{\phi}_{1:D-1}$, we actually sort the hypersurfaces at point $\myvecsymb{\phi}_{1:D-1}$ according to their magnitude. The key observation in our algorithm, is that due to the continuity of the hypersurfaces in the $\Phi^{D-1}$ hypercube, we expect that in an area ``around'' $\myvecsymb{\phi}_{1:D-1}$ the hypersurfaces will retain their magnitude-sorting. So we expect the formation of cells in the $\Phi^{D-1}$ hypercube, within which the magnitude-sorting of the hypersurfaces will remain unaltered, irrespectively of whether the magnitude of each hypersurface changes. Moreover, even if the sorting of the hypersurfaces changes at some point around $\myvecsymb{\phi}_{1:D-1}$ it is possible that the $\set{I}$ does not change. So we expect the formation of regions in the $\Phi^{D-1}$ hypercube which expand over more than one cells and within which the $\set{I}$-set remains unaltered, even if the sorting of the hypersurfaces changes. If we can efficiently determine all these cells (or even better regions) and obtain the corresponding $\set{I}$-sets, then the set of all candidate index-sets may be significantly smaller than the set of all $\binom{N}{K}$ possible index-sets. Once all the candidate $\set{I}$-sets have been collected, $\set{I}_{\text{opt}}$ and $\myvec{x}_{\text{opt}}$ will be determined through exhaustive search among the candidate sets.

In \eqref{observation_for_given_phi_is_rank_1} we observed that at a given point $\myvecsymb{\phi}_{1:D-1}$ the maximization problem resembles the \hyperref[section:rank_1_case]{rank-$1$ case} and consequently, the $\set{I}$-set at $\myvecsymb{\phi}_{1:D-1}$ consists of the indices of the $K$ largest elements of $|\mymatrix{V}\myvec{c}(\myvecsymb{\phi}_{1:D-1})|$.
Motivated by this observation, we define a \textit{labeling function} $\labelfunc{\cdot}$ that maps a point $\myvecsymb{\phi}_{1:D-1}$ to an index-set
\begin{equation}
\labelfunc{\mymatrix{V}_{N\times D};\myvecsymb{\phi}_{1:D-1}}\stackrel{\vartriangle}{=}\arg\max_{\set{I}}{\sum_{i \in \set{I}}{|\big(\mymatrix{V} \myvec{c}(\myvecsymb{\phi}_{1:D-1}) \big)_i|}}.
\end{equation}
Then, each point $\myvecsymb{\phi}_{1:D-1} \in \Phi^{D-1}$ is mapped to a candidate index-set and the optimal index-set $\set{I}_{\text{opt}}$ belongs to
\begin{equation}
 \grandset{I}_{\text{tot}}(\mymatrix{V}_{N \times D}) 
 \stackrel{\vartriangle}{=} \hspace{-.5cm}\bigcup_{\myvecsymb{\phi}_{1:D-1} \in \Phi^{D-1}} \hspace{-.5cm}\big\{\labelfunc{\mymatrix{V}_{N\times D}; \; \myvecsymb{\phi}_{1:D-1}}\big\}.
\end{equation}
In the following, we
{\it (i)} show that the total number of candidate index-sets is $|\grandset{I}(\mymatrix{V}_{N \times D})| \le \displaystyle\sum_{d=0}^{\left\lfloor \frac{D-1}{2} \right\rfloor}\binom{N}{D-2d}\binom{D-2d}{\left\lfloor\frac{D}{2}\right\rfloor-d}2^{D-1-2d}=\bigonotation{N^D}$ and {\it(ii)} develop an algorithm for the construction of $\grandset{I}(\mymatrix{V}_{N \times D})$ with complexity $\bigonotation{N^{D+1}}$.

The labeling function is based on pair-wise comparisons of the elements of $\mymatrix{V} \myvec{c}(\myvecsymb{\phi}_{1:D-1})$ while
each element of $|\mymatrix{V}_{N\times D}\myvec{c}(\myvecsymb{\phi}_{1:D-1})|$ is a continuous function of $\myvecsymb{\phi}_{1:D-1}$, a $D$-dimensional hypersurface, and any point $\myvecsymb{\phi}_{1:D-1}$ is mapped to an index-set $\set{I}$ which is determined by comparing the magnitudes of these hypersurfaces at $\myvecsymb{\phi}_{1:D-1}$. Due to the continuity of hypersurfaces, the index-set $\set{I}$ does not change in the ``neighborhood'' of $\myvecsymb{\phi}_{1:D-1}$.
A necessary condition for the $\set{I}$ set to change is two of the hypersurfaces to change their magnitude ordering.
The switching occurs at the intersection of two hypersurfaces
where we have
$\big|\big(\mymatrix{V} \myvec{c}(\myvecsymb{\phi}_{1:D-1}) \big)_i\big| = \big|\big(\mymatrix{V} \myvec{c}(\myvecsymb{\phi}_{1:D-1}) \big)_j\big|, \quad i\neq j,$
which yields
\begin{equation}
\phi_1  = \tan^{-1}{\left(  -\frac{\transp{({\bf V}_{i,2:D}\mp {\bf V}_{j,2:D})} \myvec{c}(\myvecsymb{\phi}_{2:D-1}) }{{\bf V}_{i,1}\mp{\bf V}_{j,1}}  \right)}.
\label{phi1_equation_for_partitioning_positive}
\end{equation}
Functions $\phi_1  = \tan^{-1}{\left(  -\frac{\transp{({\bf V}_{i,2:D}- {\bf V}_{j,2:D})} \myvec{c}(\myvecsymb{\phi}_{2:D-1}) }{{\bf V}_{i,1}-{\bf V}_{j,1}}  \right)}$ and $\phi_1  = \tan^{-1}{\left(  -\frac{\transp{({\bf V}_{i,2:D}+ {\bf V}_{j,2:D})} \myvec{c}(\myvecsymb{\phi}_{2:D-1}) }{{\bf V}_{i,1}+{\bf V}_{j,1}}  \right)}$ determine $(D-1)$-dimensional hypersurfaces $S({\bf V}_{i,:}\,;\,{\bf V}_{j,:})$
and $S({\bf V}_{i,:}\, ;\, -{\bf V}_{j,:})$, respectively.
Each hypersurface partitions $\Phi^{D-1}$ into two regions.

For convenience, in the following we use a pair $\{i,j\}$ to denote the rows of matrix $\mymatrix{V}_{N \times D}$, that originate hypersurface $\hypersurface{\bf V}{i}{j}$.
Moreover, we allow $i$ and $j$ to be negative in order to encapsulate the information about the sign with which each row participates in the generation of hypersurface $S$, i.e.
\begin{flalign}
 \{i,\, j\} \mapsto \hypersurface{\bf V}{i}{j}, 
\end{flalign}
where $i,j \in \{-N,\hdots, -1,1,\hdots,N\}, \; |i|\neq |j|$.

Let $\{i_1,i_2,\ldots,i_D\}\subset\{1,2,\ldots,N\}$ where $\{i_1,i_2,\ldots,i_D\}$ is one among the $\binom{N}{D}$ size-$D$ subsets of $\{1,2,\ldots,N\}$.
Then, by keeping $i_1$ fixed (where $i_1$ is arbitrarily selected, say $i_1$ is the minimum among $i_1,i_2,\ldots,i_D$) and assigning signs to $i_2,i_3,\ldots,i_D$ we can generate $2^{D-1}$ sets of the form $\{i_1,\pm i_2,\ldots,\pm i_D\}$.
Hence, we can create totally $\binom{N}{D}2^{D-1}$ such sets which we call $J_1,J_2,\ldots,J_{\binom{N}{D}2^{D-1}}$.
We can show (the proof is omitted due to lack of space) that, for any $l=1,2,\ldots,\binom{N}{D}2^{D-1}$, the $\binom{D}{2}$ hypersurfaces $\hypersurface{\bf V}{i}{j}$ that we obtain for $\{i,j\}\subset J_l$ have a single common intersection point $\hat{\myvecsymb{\phi}}(J_l)$ which ``leads'' at most $\binom{D}{\left\lfloor\frac{D}{2}\right\rfloor}$ cells.
Each such cell is associated with an index-set in the sense that $\labelfunc{\mymatrix{V}_{N \times D};\myvecsymb{\phi}_{1:D-1}}$ is maintained for all $\myvecsymb{\phi}_{1:D-1}$ in the cell.
In other words, the $\set{I}$-set associated with all points $\myvecsymb{\phi}_{1:D-1}$  in the interior of the cell is the same as the $\set{I}$-set at the leading vertex.
In fact, the actual sorting of $|\mymatrix{V}\myvec{c}(\phi_{1:D-1})|$ for all points in the interior of the cell is the same as the sorting at the leading vertex and the $\set{I}$-set may characterize a greater area that includes many cells.
In addition, we can show that examination of all such cells is sufficient for the computation of all index-sets that appear in the partition of $\Phi^{D-1}$ and have a leading vertex.

We collect all index-sets into
$\grandset{I}(\mymatrix{V}_{N \times D})$
and observe that $ \grandset{I}(\mymatrix{V}_{N \times D})$ can only be a subset of the set of all possible ${\binom{N}{K}}$ index-sets.
In addition, since the cells are defined by a leading vertex, we conclude that there are at most $\binom{N}{D}\binom{D}{\left\lfloor\frac{D}{2}\right\rfloor}2^{D-1}$ cells.
We finally note that there exist cells that are not associated with an intersection-vertex.
We can show that such cells can be ignored unless they are defined when $\phi_{D-2}=\frac{\pi}{2}$.
In the latter case, we just have to identify the cells that are determined by the reduced-size matrix $\mymatrix{V}_{N \times (D-2)}$ over the hypercube $\Phi^{D-3}$.
Hence, $\grandset{I}_\text{tot}(\mymatrix{V}_{N \times D}) = \grandset{I}(\mymatrix{V}_{N \times D}) \cup \grandset{I}_\text{tot}(\mymatrix{V}_{N \times (D-2)})$ and, by induction,
\begin{flalign}
 \grandset{I}_\text{tot}(\mymatrix{V}_{N \times d}) = \grandset{I}(\mymatrix{V}_{N \times d}) \cup \grandset{I}_\text{tot}(\mymatrix{V}_{N \times (d-2)}),\;3\leq d\leq D \nonumber,
\end{flalign}
which implies that 
\begin{flalign}
  &\grandset{I}_\text{tot}(\mymatrix{V}_{N \times D}) \nonumber\\
&= \grandset{I}(\mymatrix{V}_{N \times D}) \cup  \grandset{I}(\mymatrix{V}_{N \times (D-2)}) \cup \hdots \cup \grandset{I}(\mymatrix{V}_{N \times (D-2 \lfloor\frac{D-1}{2} \rfloor)})\nonumber\\
&= \bigcup_{d=0}^{\lfloor\frac{D-1}{2} \rfloor}\grandset{I}(\mymatrix{V}_{N \times (D-2d)}) \label{total_index_set_as_join}.
\end{flalign}
As a result, the cardinality of $\grandset{I}_\text{tot}(\mymatrix{V}_{N \times D})$ is
{\footnotesize
\begin{flalign}
 &|\grandset{I}_\text{tot}(\mymatrix{V}_{N \times D}) | \nonumber\\
&\le |\grandset{I}(\mymatrix{V}_{N \times D})|+|\grandset{I}(\mymatrix{V}_{N \times (D-2)})|+\hdots+|\grandset{I}(\mymatrix{V}_{N \times (D-2\lfloor \frac{D-1}{2} \rfloor)})|\nonumber\\
&\le \binom{N}{D}\binom{D}{\left\lfloor\frac{D}{2}\right\rfloor}2^{D-1} + \binom{N}{D-2}\binom{D-2}{\left\lfloor\frac{D}{2}\right\rfloor-1}2^{D-3} + \hdots\nonumber\\
&+ \binom{N}{D-2\lfloor \frac{D-1}{2} \rfloor}\binom{D-2\lfloor \frac{D-1}{2} \rfloor}{\left\lfloor\frac{D}{2}\right\rfloor-\left\lfloor\frac{D-1}{2}\right\rfloor}2^{D-1-2\lfloor \frac{D-1}{2} \rfloor}\nonumber\\
& = \sum_{d=0}^{\left\lfloor \frac{D-1}{2} \right\rfloor}\binom{N}{D-2d}\binom{D-2d}{\left\lfloor\frac{D}{2}\right\rfloor-d}2^{D-1-2d}=\bigonotation{N^D}.
\end{flalign}
}%
It remains to show how $\grandset{I}(\mymatrix{V}_{N \times D})$ is constructed.

As already mentioned, there are in total $\binom{N}{D}2^{D-1}$ intersection points which can all be blindly examined.
For any $l=1,2,\ldots,\binom{N}{D}2^{D-1}$, the cell leading vertex $\hat{\myvecsymb{\phi}}(J_l)$ is computed efficiently as the intersection of $D-1$ hypersurfaces,
i.e. the unique solution of
\begin{flalign}
\begin{pmatrix}
\begin{array}{l@{}l}
  \mymatrix{V}_{ i_1,\, 1:D}&\mp\mymatrix{V}_{ i_2,\, 1:D}\\
  &\vdots\\
  \mymatrix{V}_{ i_1,\, 1:D}&\mp\mymatrix{V}_{ i_D,\, 1:D}
\end{array}
\end{pmatrix}
%
\myvec{c}(\myvecsymb{\phi}_{1:D-1})
= \myvec{0}_{(D-1)\times 1},
\label{find_phi_with_svd}
\end{flalign}
which is obtained in time $\bigonotation{1}$ with respect to $N$.
After $\hat{\myvecsymb{\phi}}(J_l)$ has been computed, we have to identify the index-sets associated with the cells that originate at $\hat{\myvecsymb{\phi}}(J_l)$ by calling the labeling function $\labelfunc{\mymatrix{V}_{N\times D}\,;\, \hat{\myvecsymb{\phi}}(J_l)}$ which marks the $K$ elements of the $\set{I}$-set, i.e. the indices of the $K$ largest elements of $|\mymatrix{V}\myvec{c}(\hat{\myvecsymb{\phi}}(J_l))|$, $l=1,2,\ldots,\binom{N}{D}2^{D-1}$.
Since $\hat{\myvecsymb{\phi}}(J_l)$ constitutes the intersection of $D$ hypersurfaces that correspond to the $D$ elements of $\hat{\myvecsymb{\phi}}(J_l)$, the corresponding values in $|\mymatrix{V}\myvec{c}(\hat{\myvecsymb{\phi}}(J_l))|$ equal each other.
If the $K$ largest values in $|\mymatrix{V}\myvec{c}(\hat{\myvecsymb{\phi}}(J_l))|$ contain $D_l$ values that correspond to the $D$ elements of $J_l$, then we can blindly examine all $\binom{D}{D_l}$ cases of index-sets to guarantee that all actual index-sets are included.
The term $\binom{D}{D_l}$ is maximized for $D_l=\left\lfloor\frac{D}{2}\right\rfloor$, hence at most $\binom{D}{\left\lfloor\frac{D}{2}\right\rfloor}$ index-sets correspond to intersection point $\hat{\myvecsymb{\phi}}(J_l)$.

The complexity to build $\grandset{I}(\mymatrix{V}_{ N\times D})$ results from the parallel examination of $\binom{N}{D}2^{D-1}$ intersection points while worst-case complexity $\bigonotation{N}+\binom{D}{\left\lfloor\frac{D}{2}\right\rfloor}$ is required at each point for the identification of the corresponding index-sets.
The term $\bigonotation{N}$ corresponds to the cost required to determine the $K$th-order element of $|\mymatrix{V}\myvec{c}(\hat{\myvecsymb{\phi}}(J_l))|$ in an unsorted array.
Consequently, the worst-case complexity to build $\grandset{I}(\mymatrix{V}_{ N\times D})$ becomes $\binom{N}{D}2^{D-1}\times\left(\bigonotation{N}+\binom{D}{\left\lfloor\frac{D}{2}\right\rfloor}\right)=\bigonotation{N^{D+1}}$.
Finally, we recall that the cardinality of $\grandset{I}_\text{tot}(\mymatrix{V}_{N \times D})$ is upper bounded by $\bigonotation{N^D}$ and conclude that the overall complexity of our algorithm for the evaluation of the sparse principal component of $\mymatrix{V}\mymatrix{V}^T$ is upper bounded by $\bigonotation{N^{D+1}}$.

\section{Conclusions}
We considered the problem of identifying the sparse principal component of a rank-deficient matrix.
We introduced auxiliary spherical variables and proved that there exists a set of candidate index-sets whose size is polynomially bounded, in terms of rank, and contains the optimal index-set, i.e. the index-set of the nonzero elements of the optimal solution.
Finally, we developed an algorithm that computes the optimal sparse principal component in polynomial time.
Our proposed algorithm stands as a constructive proof that the computation of the sparse principal component of a rank-deficient matrix is a polynomially solvable problem.

\end{document}